\def   \ni {\noindent}

\def   \sax {BeppoSAX} 
\def   \ssk {\vskip  5truept}

\def   \bsk {\vskip 15truept}
 
\def   \newpage {\vfill\eject}
\def   \newline {\hfil\break}

\documentstyle[epsfig]{article}
\begin{document}

\hsize 5truein
\vsize 8truein
\font\abstract=cmr8
\font\keywords=cmr8
\font\caption=cmr8
\font\references=cmr8
\font\text=cmr10
\font\affiliation=cmssi10
\font\author=cmss10
\font\mc=cmss8
\font\title=cmssbx10 scaled\magstep2
\font\alcit=cmti7 scaled\magstephalf
\font\alcin=cmr6 
\font\ita=cmti8
\font\mma=cmr8
\def\ref{\par\noindent\hangindent 15pt}
\null
%\vskip 3.0truecm
%\baselineskip = 12pt

% ------ beginning of font "title" ------

\title{\ni Bright Radio Galaxies with \sax}

% beginning of font "author and affiliation"
\bsk \bsk
\author{\ni P.~Grandi $^{1}$, L.~Maraschi$^{2}$, M. Guainazzi$^3$, 
F. Haardt$^{4}$, G. Matt$^5$, E. Massaro$^6$, C. M. Urry$^7$,
L. Bassani$^8$, P. Giommi$^9$}

\ssk
\affiliation{1) IAS/CNR, Area di Ricerca di Tor Vergata, Roma I-00133, Italy}

\affiliation{2) Osservatorio Astronomico di Brera, Milano, Italy} 

\affiliation{3) SSD/ESA, ESTEC, Postbus  299, 2200 AG Noordwijk, The Netherlands}

\affiliation{4) Universita' dell'Insubria/Polo  di Como, Como, Italy}  

\affiliation{5) Universita' degli Studi "Roma 3", Roma, Italy}

\affiliation{6) Istituto Astronomico, Universita' di Roma, Via Lancisi 29 Roma, Italy}

\affiliation{7) StscI, Baltimore, MD, USA}  

\affiliation{8) ITESRE/CNR, Bologna, Italy}

%%\affiliation{9) Osservatorio Astronomico di Acetri, Firenze, Italy}

%\affiliation{10)Universita' di Ferrara, Ferrara, Italy}

\affiliation{9) BeppoSAX SDC, c/o Nuova Telespazio, Roma, Italy}

%\affiliation{12)IFCAI/CNR, Palermo, Italy}
\bsk
\baselineskip = 12pt

% beginning of font "abstract and keywords"
\abstract{ABSTRACT \ni
Although  a strong similarity between radio 
galaxies and Seyferts has been pointed out with ASCA, some radio galaxies 
do not seem to fit very well the cold thin accretion disk 
model proposed for radio quiet AGN.
The \sax~ observations of 3C390.3 and Centaurus A 
show that cold material responsible  for the photon reprocessing 
is present in these sources but not necessarily near to the primary X-ray source.
The featureless spectrum of the broad line radio galaxy 3C111
indicates the absence of cold material 
near the X-ray source or, alternatively, the presence of a strong 
jet X-ray component.
}                                                    
\bsk
\baselineskip = 12pt

\bsk
\baselineskip = 12pt

% beginning of font "text"

\text{\ni 1. INTRODUCTION
\ssk
\ni     
Radio Galaxies have been poorly investigated in the past, as shown by 
scarce and sometime contradictory results from previous satellites.
The bright radio galaxies, in particular the Fanaroff-Riley (FR) II,
have been often treated as Seyfert galaxies
and included in sample of radio quiet AGN without taking into account their 
radio loudness.  It was implicitly assumed that 
they share the same nuclear properties of the Seyfert galaxies, in 
agreement with the unified scheme for AGN
that considers the Broad Line Radio 
Galaxies (BLRG) and the Narrow Line Radio galaxies (NLRG) the radio 
loud counterparts of the Seyfert 1 and Seyfert 2, respectively.
The detection of broad iron lines  
in several BLRG with ASCA seemed to confirm this picture;
the large intrinsic width ($\sigma> 0.6$ keV) observed
in 3C120, 3C382, 3C109  (Allen et al. 1997, Grandi et al. 1997, Reynolds 1997)
suggested that the line is produced 
in the inner regions of the accretion disk as it is thought to be the case 
in Seyferts. The detection of a strong UV bump (Maraschi et al. 1991) 
and the indication of  a possible soft excess (Grandi et al. 1997)
in 3C120 further strengthened this hypothesis.

The \sax~ observations of bright radio galaxies reported here show that 
\newpage\noindent
the similarity between radio quiet and radio loud AGN
is not a foregone conclusion.

At least two questions needed to be addressed: 1) are the 
accretion processes the same in radio galaxies and Seyferts?
2) Is really the X-ray jet component negligible in 
the radio galaxies?  

\ssk
\ni 2. BeppoSAX OBSERVATIONS 
\ssk

%BeppoSAX dedicated a Core Program to the Bright Radio Galaxies with 
%observed flux $F_{\rm 2-10 keV}>10^{-11}$ erg cm$^{-2}$ s$^{-1}$ and
%radio loudness RL=log(F$_{5 GHz}/F_{V}>2$.
%The aim of this project is to perform a detailed spectral study 
%in the 0.1-150 keV using long ($\sim 100$ ksec) exposure time.
%The sources observed so far are 3C390.3, Centaurus A (Cen A) and 3C111.

3C390.3 is a BLRG 
($L^{\it unabs.}_{\it 2-10 keV}\sim 3\times10^{44}$ erg sec$^{-1}$) 
characterized by a FRII morphology with a core showing superluminal motion. 
It is characterized by a weak blue bump and double peaked emission lines 
in the optical and UV bands.
%\sax~showed for the first time that 
%both the K$_{\alpha}$ iron line and a strong reflection
%component are present in this source.
The \sax~spectrum is well fitted by a power law
$\Gamma=1.80^{+0.05}_{-0.04}$ reflected at high energies by material with 
a fairly large covering factor ($\Omega/2\pi=0.9^{+0.5}_{-0.3}$).
The column density ($N_H\sim 10^{23}$ cm$^{-2}$) 
is larger than the Galactic one and probably 
variable on time scale of years. No soft X-ray excess was detected.
The iron line at 6.4 keV is intrinsically narrow ($\sigma=73^{+270}_{-73}$ 
eV) and has an equivalent width of $\sim 140$ eV 
(see Grandi et al. 1998a for more details). 
 
Although these features are typical of Seyfert galaxies
the absence of a X-ray soft excess, the presence of a weak blue bump 
and of a narrow iron line  are difficult to 
reconcile with a Seyfert like model that assumes a geometrically thin 
accretion disk and an active corona above it (Haardt $\&$ Maraschi 1991,1993).
A completely hot inner flow (Shapiro, Lightman Eardley 1976, Narayan 
Mahadevan and  Quataert 1998), surrounded  by cold material 
in shape of a warped disk or a thick dusty torus at parsec distances
is a more plausible description. A hot inner region can explain the lack 
of the soft excess and the cold external material 
the weak blue bump, the narrow iron line and the reflection.
%
%The absence of short term variability in the intensity of the 
%Fe Line recently reported by Wozniak et al. 1998 seem 
%to support this picture.

%The possibility that cold material can be located far away from 
%the X-ray primary  source is also supported by the \sax~observations 
%of Cen A.

Cen~A is a well-known NLRG ($L^{\it unabs.}_{\it 2-10 keV}\sim 10^{43}$ 
erg sec$^{-1}$) which has
been extensively studied  over the whole X-ray and 
$\gamma$-ray range (Turner et al. 1997 and refs therein).
It was observed by \sax~ twice on 1997 February 20-12 and 1998 January 6-7
for 35 and 75 ksec, respectively.
The average source flux increased between the two observations, being
brighter by about a factor 1.3 in 1998.
Given the complexity of this source we initially focused on 
the nuclear point-like emission.
We studied the 1.5-150 keV spectrum (MECS and PDS data), fixing 
the contributions of the extended components 
(see Grandi et al. 1998b).
At both epochs, the nuclear point-like emission is well fitted
with a similar strongly absorbed ($N_H\sim10^{23}$ cm$^{-2}$)
power law ($\Gamma\sim 1.7$) with an exponential cutoff at high energies
(E$_{cut}> 200$~keV). The 1997 data need the contribution of a weak 
reflection component ($\Omega/2\pi\sim 0.2$), which should be
considered with caution because of 
intercalibration uncertainties ($10\%$) still present between MECS and PDS.
%We did not detect any spectral variation of the continuum between the 
%two observations.
A significant flux variation of the iron line was observed between
the two observations. The flux of the line and of the continuum
changes in the opposite sense. The line is more intense at the first epoch,
when the nuclear source is at the lower intensity level
(see figure 2 in Grandi et al. 1998b).
The implied delay between the continuum and line variations strongly suggests
that the cold material responsible for the iron line production
can not be located (as in the case of 3C390.3) very near to the primary 
X-ray source. 

3C111 is a BLRG 
($L^{\it unabs.}_{\it 2-10 keV}\sim 3\times10^{44}$ erg sec$^{-1}$) 
with radio properties similar to 3C390.3.
%It is a FRII and shows a superluminal motion ($\beta_{app}=3.4$).
It is a strongly absorbed source, being located behind the galactic dark 
cloud, Taurus B. 

BeppoSAX observed this source for about 100 ksec.
The spectrum is well fitted by a simple power law with a quite hard
spectral index ($\Gamma\sim1.7$). Neither reflection component nor iron 
line are required by the data. We could only estimate an upper limit for 
the 6.4 keV fluorescence feature (EW$< 66 eV$) and for the amount of 
reflection (R$<0.13$) (Grandi et al. in preparation).
In agreement with the ASCA results (Reynolds et al. 1998), 
the low energy absorption requires a column density 
($N_H=10(\pm0.1)\times 10^{21}$ cm$^{-2}$) which is larger than
that estimated in the direction of Taurus B 
($N_H\sim 3\times 10^{21}$ cm$^{-2}$; Ungerer et al. 1985, 
Elvis Lockman and Wilkes 1989).
The interpretation of the data is not immediate.
At least two possible scenarios can be prospected. 
1) The absorption due to the Taurus B cloud
is underestimated because the current 
measures of N$_H$ do not take into account sub-parsec molecular
inhomogeneities (see Reynolds et al. (1998)  for an accurate discussion).
In this case, the X-ray photons are absorbed in our Galaxy,
there is not intrinsic absorption in 3C111 and the spectrum is 
blazar-like. 2) Intrinsic photons absorption occurs in the source and, as 
suggested by the superluminal motion observed at radio 
frequencies, we are directly seeing the nuclear region.
The lack of signatures of cold material, 
like the iron line and the reflection component,  
might indicate the presence of an accretion flow (or a considerable part of it)
too hot to produce typical Seyfert 1 features. 
Note that another BLRG, Pictor A, is characterized by a similar X-ray spectrum 
(Padovani  et al. 1998, Eracleous and Halpern 1998).

\ssk
\ni 3. CONCLUSIONS 
\ssk
\ni 
The similarity between radio-quiet and radio loud AGN
has still to be confirmed.
Recent observations of radio galaxies have shown that there is 
not a unique type of X-ray spectrum (see Figure 1)

\begin{figure}[th]
\centerline{\psfig{figure=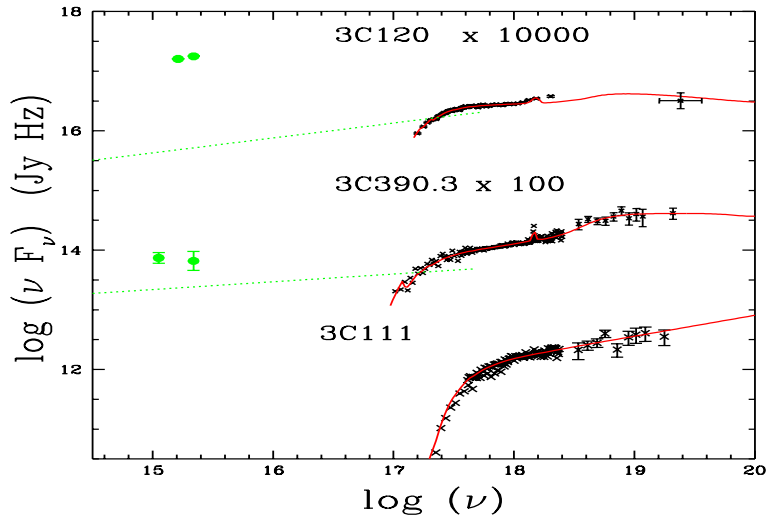,height=7.1cm,width=10.5cm}}
\caption{FIGURE 1. Spectral energy distributions of three 
types of BLRG. 3C120 shows a Seyfert-like spectrum with strong 
blue bump, broad iron line and probably reflection.
3C390.3 is characterized by a weak blue bump and
a narrow iron line. 3C111 shows neither 
the iron line nor the Compton hump (BeppoSAX data).
The UV points and the extrapolated X-ray power laws (dotted blue lines) 
refer to simultaneosly IUE-EXOSAT 
(3C120: Maraschi et al. 1991) and IUE-ROSAT/PSPC 
(3C390.3: Walter et al. 1991, A$\&$A, 285, 119) observations.
The ASCA and OSSE data of 3C120 are from Grandi et al. 1997, the 
\sax~data of 3C390.3 from Grandi et al. 1998a.}
\end{figure}

Some sources (3C120, 3C109, 3C382) fit very well the Seyfert-like model.
The broad iron line observed by ASCA can be easily explained 
assuming cold material (a geometrically thin disk) rapidly 
rotating very near the black hole.
In this case the X-ray primary photons are assumed to be produced by an 
active corona embedding the inner region of the accretion disk.

In other cases (3C390.3 and Cen A),  a geometry 
which assumes cold material far away from the X-ray primary source
seems to provide a better explanation of the observations.
In particular for 3C390.3, it is possible that the 
X-ray continuum  is produced by a hot inner flow while the reprocessed 
radiation come from an outer region shaped as a warped disk and/or a torus.

Some objects (3C111, Pictor A) do not show the iron line and, 
in the case of 3C111, the Compton hump is also absent.
The beamed jet component could be important in these sources 
and obscure the Seyfert-like emission.
Alternatively,  the accretion flow could  be
hot, maybe in a two temperature configuration (as that suggested for 
the inner region of 3C390.3) .
}

\bsk
\baselineskip = 10.5pt

% beginning of font "references"

{\references \ni REFERENCES
\ssk
\ref Allen S. W. et al. , 1997, MNRAS, 286, 765 
\ref Elvis M., Lockman, F. J., Wilkes, B. J., 1989, AJ 97, 777
\ref Eracleous M.  Halpern J. P., 1998 ApJ in press.
\ref Grandi P., et al. 1997, ApJ, 487, 636 
\ref Grandi P., et al. 1998a, A$\&$A, in press (astro--ph/9810450)
\ref Grandi P., et al. 1998b , Proceeding of the 32nd Cospar Meeting, 
Nagoya, Japan 
\ref Haardt F., Maraschi L, 1991, ApJ, 380, L51
\ref Haardt F., Maraschi L., 1993, ApJ, 413, 507
\ref Maraschi L., et al., 1991, APJJ 368, 138
\ref Narayan R., Mahadevan R. and Quataert E. 1998, {\it The
        Theory of Black Hole Accretion Disk}, eds M.A. Abramowicz,
        G. Bjornsson, J.E. Pringle, in press (astro--ph/9803141)
\ref Padovani P. et al. 1998 MNRAS in press 
\ref Reynolds C. S., 1997, MNRAS 286, 513
\ref Reynolds C. S., Iwasawa K., Crawford C. S. Fabian A. C., 1998
MNRAS in press
\ref  Shapiro S.L., Lightman A.P., Eardley D.M., 1976, ApJ, 204, 187
\ref Turner T. J., et al., 1997, ApJ, 475, 118 
\ref Ungerer et al., 1985,  A$\&$A, 146, 123 
\ref Wozniak P. R., et al. 1988, MNRAS, 299, 449 
}                      

\end{document}